\documentstyle[12pt]{article}
\def\be{\begin{equation}}
\def\ee{\end{equation}}
\def\ba{\begin{eqnarray}}
\def\ea{\end{eqnarray}}

\begin{document}

\begin{titlepage}
\null\vspace{-62pt}
\begin{flushright} hep-ph/9802356 \\
IC/98/12 \\
February 1998
\end{flushright}
\vspace{0.2in}

\centerline{{\large \bf Hierarchical Quark Mass Matrices}}

\vspace{0.5in}
\centerline{
Andrija Ra\v{s}in }
\vspace{0.2in}
\centerline{\it High Energy Section}
\centerline{\it International Center for Theoretical Physics}
\centerline{\it Strada Costiera 11}
\centerline{\it 34014 Trieste}
\centerline{\it Italy}
\vspace{.7in}
\baselineskip=19pt

\centerline{\bf Abstract}
\begin{quotation}

I define a set of conditions that the most general hierarchical Yukawa
mass matrices have to satisfy so that the leading rotations in the
diagonalization matrix are a pair of (2,3) and (1,2) rotations. In
addition to Fritzsch structures, examples of such hierarchical structures
include also matrices with (1,3) elements of the same order or even much
larger than the (1,2) elements. Such matrices can be obtained in the
framework of a flavor theory.
 
To leading order, the values of the angle in the (2,3) plane ($s_{23}$) 
and the angle in the (1,2) plane ($s_{12}$) do not depend on the order in
which they are taken when diagonalizing.  We find that any of the
Cabbibo-Kobayashi-Maskawa matrix parametrizations that consists of at
least one (1,2) and one (2,3) rotation may be suitable. In the particular
case when the $s_{13}$ diagonalization angles are sufficiently small
compared to the product $s_{12}s_{23}$, two special CKM parametrizations
emerge: the $R_{12}R_{23}R_{12}$ parametrization follows with $s_{23}$
taken before the $s_{12}$ rotation, and vice versa for the
$R_{23}R_{12}R_{23}$ parametrization. 

\vspace{0.4in}

%PACS numbers: xx.xx

\end{quotation}
\end{titlepage}

\baselineskip=19pt

\vspace{0.2in}

{\bf A. Introduction}\hspace{0.5cm}

A hierarchical structure of the Yukawa matrix is the most widely used
structure. It can follow, for example, from flavor theories with either
abelian or nonabelian symmetries. In theories with abelian symmetries the
hierarchy is obtained by assigning different charges to different
families\cite{frog79}. Families that have a larger
charge will have a higher power of the flavor symmetry breaking parameter and
thus will have a smaller Yukawa coupling. In theories with 
nonabelian symmetries, the  hierarchy in the couplings is a reflection of
the hierarchy in symmetry breaking scales\cite{bere83}.
Hierarchies can also be generated radiatively where the small numbers
originate in the loop factors\cite{wein72}.
General hierarchical structures, but only texture zeroes, 
have been studied before
\cite{matu82,ramo93,bran94}. Another very popular structure, which we do 
not consider here, is a democratic one\cite{hara78} where the elements are
all of order one and sufficiently close to
each other so that only one eigenvalue is large. 
Other structures may combine hierarchy and democracy\cite{ynir94,rasi94}.

In this paper I give a set of conditions that define the most general
hierarchical matrix, with the condition that leading 
rotations in the diagonalization matrix are a pair of $s_{12}$ and
$s_{23}$ rotations. 
%%% addition
In what follows I will assume that there are no
large accidental cancellations between the up and down mixing
angles, so that the hierarchies in both up and down sector
are of the same order or smaller than the 
corresponding observed quark masses and mixings.
%%% end addition
If the up and down quark mass matrices are
hierarchical, at least one of them, if not both
\footnote{There may be a simpler up or down matrix, {\it i.e.} with
mixings between only two generations, which is diagonalized with only one
rotation.}, must fall into the above category. 
In addition to the well known Fritzsch structures,
the hierarchy conditions permit structures which may have a large (1,3)
element. 

Next, I discuss possible parametrizations of the 
Cabibbo-Kobayashi-Maskawa (CKM) matrix\cite{cabi63,koba73} 
that emerge
from hierarchical mass matrices. I will use some recently obtained {\it
exact} results about diagonalizing $3\times3$ matrices \cite{rasi97}, in
order to control the corrections involving small terms (for example
the rotation angles in the (1,3) plane, $s_{13}$). The basic result is
that any CKM parametrization that has at least one (2,3) rotation and
one (1,2) rotation is practical. 
Which one of the parametrizations should be used will at the end depend on
the flavor theory, {\it i.e.} the explicit structure of the Yukawa matrices.
If the theory has a prediction, {\it e.q.} if some of the diagonalizing
angles can be expressed in terms of quark masses, it might be obvious
in one parametrization but not in another.

A particular example is the case when the $s_{13}$ angles are small
compared to the $s_{12}s_{23}$ product. Two parametrizations emerge as
winners: the ``$R_{12} R_{23} R_{12}$" parametrization (proposed
by Dimopoulos, Hall and Raby\cite{dimo92}(see also \cite{hall93,
ande93,barb96});
it was recently proposed as ``standard" by Fritzsch and Xing
\cite{frit97}) and the ``$R_{23} R_{12} R_{23}$" parametrization (the
original
Kobayashi-Maskawa parametrization\cite{koba73}). 
It will depend on the underlying flavor theory
that predicted the hierarchical structures
which one of these two parametrizations should be used.
If one has precise predictions for the
$s_{12}$ rotations in terms of quark masses 
one should use the first parametrization.
Conversely, if one can predict more precisely the $s_{23}$ rotations in
terms of the quark masses, one should use the second parametrization.

In the next section we review some of the notation and results about
diagonalizing
quark mass matrices from reference \cite{rasi97}. In Section C, we define
the hierarchical structures of Yukawa matrices and list some illustrative
examples. Interesting structures emerge beyond the more familiar
Fritzsch type ones. Then we turn to the question of which CKM parametrization
is most practical to use for the hierarchical structures.
First, in Section D we show that the values of $s_{23}$ and $s_{12}$ do not depend
on the order in which the rotations are taken when diagonalizing the mass matrices. 
Using this result we compare various CKM parametrizations
for hierarchical structures in Section E. We present examples of
predictions with particular CKM parametrizations and conclude 
in Section F.

\vspace{0.2in}

{\bf B. Diagonalizing Quark Mass Matrices}\hspace{0.5cm} 

Following the notation of \cite{rasi97}, we denote the Yukawa matrices as
\be
u^c {\bf h}^u Q  +  
d^c {\bf h}^d Q  \, .
\ee
Each of the matrices ${\bf h}^{u,d}$ is diagonalized by a
biunitary transformation

\be
{\bf m} 
= 
{\bf S}^\dagger 
{\bf h}
{\bf R} \, .
\ee
The matrices
${\bf S}$ and ${\bf R}$  diagonalize the following products of ${\bf h}$
\be
{\bf m^2} = {\bf S}^\dagger {\bf h} {\bf h}^\dagger {\bf S} \, \, , \, \,
{\bf m^2} = {\bf R}^\dagger {\bf h}^\dagger {\bf h} {\bf R} \, .
\label{vdef}
\ee
The CKM matrix is given by
\be
{\bf V} = {\bf R}^{u\dagger} {\bf R}^d \, .
\ee

Let us neglect phases for the moment \footnote{A complete diagonalization
of the general case with complex phases was given in \cite{rasi97}.}, we
will discuss them later in the text. Then the matrix 
${\bf h}^\dagger {\bf h}$ is of the form

\be
{\bf h}^\dagger {\bf h} =
\left(
\begin{array}{ccc}
\lambda_{11} & \lambda_{12} & \lambda_{13} \\
\lambda_{12} & \lambda_{22} & \lambda_{23} \\
\lambda_{13} & \lambda_{23} & \lambda_{33}
\end{array}
\right)  \, ,
\ee
%%% begin addition
where we assume all the elements to be non-negative (i.e. any
negative signs are absorbed with the phases which are discussed
later).
%%% end addition
Since this matrix is hermitean, the eigenvalues $\lambda_i$ are real and
nonnegative. They can be found as the solution of the 
cubic equation in $\lambda$

\ba
\det({\bf h}^\dagger {\bf h} - \lambda {\bf 1}) & = & 
 -\lambda^3 + \lambda^2(\lambda_{11} +\lambda_{22} +\lambda_{33})
- \lambda(\lambda_{11}\lambda_{22}+\lambda_{11}\lambda_{33}+
\lambda_{22}\lambda_{33}
-\lambda^2_{23}-\lambda^2_{13}-\lambda^2_{12}) 
\nonumber\\
& & 
+\lambda_{11}(\lambda_{22}\lambda_{33}-\lambda^2_{23})
-\lambda_{12}(\lambda_{12}\lambda_{33}-\lambda_{13}\lambda_{23})
+\lambda_{13}(\lambda_{12}\lambda_{23}-\lambda_{13}\lambda_{22}) = 0 \, .
\label{cubic}
\ea

The diagonalizing matrix ${\bf R}$ is a product of three plane rotations.
It is
completely arbitrary which three rotations we pick; the only requirement is that
two successive rotations are not in the same plane, since they can be trivially
combined into one rotation. 

We can write the three rotation angles in terms of the eigenvalues $\lambda_i$
and matrix elements $\lambda_{ij}$. Let us show the procedure for a choice of
rotations

\be 
{\bf R} = {\bf R}_{23} {\bf R}_{13} {\bf R}_{12} \, .
\label{leftv}
\ee
From (\ref{vdef})

\be
\left(
\begin{array}{ccc}
\lambda_1 & 0 & 0 \\
0 & \lambda_2 & 0 \\
0 & 0 & \lambda_3
\end{array}
\right)
=
{\bf R}^T_{12} {\bf R}^T_{13} {\bf R}^T_{23}
\left(
\begin{array}{ccc}
\lambda_{11} & \lambda_{12} & \lambda_{13} \\
\lambda_{12} & \lambda_{22} & \lambda_{23} \\
\lambda_{13} & \lambda_{23} & \lambda_{33}
\end{array}
\right) 
{\bf R}_{23} {\bf R}_{13} {\bf R}_{12} \, .
\ee
We can rewrite the above equation
\be
\left(
\begin{array}{ccc}
\lambda_{11} & \lambda_{12} & \lambda_{13} \\
\lambda_{12} & \lambda_{22} & \lambda_{23} \\
\lambda_{13} & \lambda_{23} & \lambda_{33}
\end{array}
\right) 
{\bf R}_{23} {\bf R}_{13} {\bf R}_{12} 
=
{\bf R}_{23} {\bf R}_{13} {\bf R}_{12}
\left(
\begin{array}{ccc}
\lambda_1 & 0 & 0 \\
0 & \lambda_2 & 0 \\
0 & 0 & \lambda_3
\end{array}
\right) \, .
\ee
Now, comparing the off diagonal elements (2,3), (1,3) and (1,2) on both
sides we can find the rotation angles
\ba
{s_{23} \over c_{23}} & = & { {(\lambda_3 -
\lambda_{11})\lambda_{23} + \lambda_{13}\lambda_{12}} 
\over {(\lambda_3-\lambda_{22})(\lambda_3-\lambda_{11})-\lambda^2_{12}}}
\, ,
\nonumber\\
{s_{13} \over c_{13}} & = & { {\lambda_{12}s_{23} + \lambda_{13}c_{23}} 
\over {\lambda_3-\lambda_{11}} }
\, ,
\nonumber\\
{s_{12} \over c_{12}} & = & { {\lambda_{12}c_{23} - \lambda_{13}s_{23} } 
\over {(\lambda_2-\lambda_{11})c_{13} +
(\lambda_{12}s_{23}+\lambda_{13}c_{23})s_{13}}} \, \, . 
\label{exact1}
\ea
where $s_{ij} \equiv \sin \theta_{ij}$
and $c_{ij} \equiv \cos \theta_{ij}$.

\vspace{0.2in}

{\bf C. Hierarchical Structures of Yukawa Matrices}\hspace{0.5cm}

In this paper I will define a {\bf hierarchical Yukawa matrix} 
as any Yukawa matrix
that has a hierarchy in the 
elements with the following conditions:

$\bullet$ There is a hierarchy in the eigenvalues $\lambda_1 << \lambda_2 <<
\lambda_3$. The cubic equation in $\lambda$ with coefficients in terms of
the eigenvalues reduces in the leading order to 
\be
(\lambda_1 - \lambda)(\lambda_2 - \lambda)(\lambda_3 - \lambda)
\approx
- \lambda^3 + \lambda^2\lambda_3
- \lambda\lambda_2\lambda_3 + \lambda_1\lambda_2\lambda_3 = 0
\, .
\label{cubicapp}
\ee

$\bullet$ The hierarchy is such that the largest element is $\lambda_{33}$.
In addition the second eigenvalue is to the leading order given in terms of
the closest neighbors of the largest element, that is $\lambda_2$ is
given in terms of $\lambda_{22}$ and $\lambda_{23}$. Comparing the
cubic equations (\ref{cubic}) and (\ref{cubicapp}) we see
\ba
\lambda_3 & \approx & \lambda_{33} \, , \nonumber\\ 
\lambda_2 & \approx &\lambda_{22} - { \lambda^2_{23} \over \lambda_3 }
\, ,
\nonumber\\
\lambda_1 & \approx & \lambda_{11} - 
{\lambda^2_{12} \over \lambda_2} -
{{\lambda_{13}(\lambda_{13}\lambda_{22}-2\lambda_{12}\lambda_{23})} \over 
{\lambda_2\lambda_3}} \, .
\ea
This is achieved with the following conditions
\ba
{\rm \bf h1)} &  \lambda_{33} >> \, \, 
{\rm all} \,\,{\rm other} \, \, \lambda_{ij} \,\, ;
\nonumber\\
{\rm \bf h2)} &  (\lambda_{22} - \lambda^2_{23} / \lambda_{33}) >> 
\lambda_{11}, \, \lambda_{12}, \, \lambda^2_{13} / \lambda_{33} \, .
\ea

$\bullet$ The (1,3) rotation needed to diagonalize such a matrix is much smaller
than the (1,2) and (2,3) rotations. 
This requirement follows from the observed CKM values, assuming there are no
accidental cancellations between the (1,3) rotations coming from diagonalizing
up and down sector. Looking at the exact results (\ref{exact1}),
for $s_{13} << s_{23}$ the condition is 
\be
{\rm \bf h3)} \, \,  \lambda_{23} >> \lambda_{13} \, \, .
\ee
Finally, for $s_{13} << s_{12}$ we need  
\ba
 {\rm \bf h4)} & 
\lambda_{13} \stackrel{<}{\sim}  \lambda_{12} s_{23} 
\, \, \, {\rm or } 
\nonumber\\
& ( \lambda_{13}  >>  \lambda_{12} s_{23} 
\, \, 
 {\rm and} 
\nonumber\\
 & (  \lambda_{22} \stackrel{<}{\sim}
\lambda_{23}^2 / \lambda_{33} \, \, \, {\rm or} 
\, \, \, ( \lambda_{12} >> \lambda_{13} \lambda_{22} / \lambda_{33} 
\, \, \, {\rm and}
\, \, \, \lambda_{12} >> \lambda_{13} \lambda_{23} / \lambda_{33} 
)))
\ea
Notice that the condition {\bf h4)} for a sufficiently small $\lambda_{22}$ does not
further constrain the element $\lambda_{13}$ (see examples III and IV 
below).  
The angles are now to leading approximation
\ba
{s_{23} \over c_{23}} & \approx & { {\lambda_{23}} \over {\lambda_3}}
\nonumber\\
{s_{13} \over c_{13}} & \approx & { {\lambda_{12}s_{23} + \lambda_{13}} 
\over {\lambda_3} }
\nonumber\\
{s_{12} \over c_{12}} & \approx & { {\lambda_{12} - \lambda_{13}s_{23} } 
\over { \lambda_2 }} \,
\label{app32}
\ea

Conditions {\bf h1)}-{\bf h4)} define the hierarchical structures. Notice that these
are the
conditions on elements of ${\bf h}^\dagger {\bf h}$, 
not on ${\bf h}$ \footnote{
An example of a Yukawa matrix ${\bf h}$ with a somewhat unusual
structure, but for which ${\bf h}^\dagger{\bf h}$ still satisfies 
conditions {\bf h1}-{\bf h4)} was given in \cite{babu}
\be
{\bf h} =
\left(
\begin{array}{ccc}
c_1 & b_1 & a_1 \\
c_2 & b_2 & a_2 \\
c_3 & b_3 & a_3 
\end{array}
\right) \, , \,
\ee
with $a_i >> b_j >> c_k$. I thank K.S. Babu for bringing this example to my
attention.}.
Conditions can be worked out for the elements of ${\bf h}$ itself.
Here we just give the mass eigenvalues 
$m_i = \sqrt{\lambda_i}$ in terms of $h_{ij}$
\ba 
m_3 & \approx & h_{33} \, \, ,
\nonumber\\ 
m_2 & \approx & |h_{22} - {{h_{23}h_{32}} \over m_3}| \, \, ,
\nonumber\\
m_1 & \approx & |h_{11} - 
{h_{12}h_{21} \over m_2} -
{{h_{13}(h_{31}h_{22}-h_{21}h_{32})-h_{31}h_{12}h_{23}} \over 
{m_2m_3}} |\, .
\ea

In what follows we will only need the actual conditions {\bf h1)}-{\bf h4)}.

Let us now show five illustrative examples of hierarchical structures, which
have certain relations between diagonalizing angles and quark masses. For 
simplicity I assume 
that the structures are themselves hermitean in the first four examples so that
conditions {\bf h1)}-{\bf h4)}
apply to elements of ${\bf h}$ itself\footnote{The only difference is that
some eigenvalues may be negative, which can be simply corrected by a sign
redefinition of the fields.}. The fifth example is an asymmetric matrix.

{\bf Example I:}\cite{frit77} 

\be
{\bf h} =
\left(
\begin{array}{ccc}
0 & \lambda_{12} & 0 \\
\lambda_{12} & 0 & \lambda_{23} \\
0 & \lambda_{23} & \lambda_{33} 
\end{array}
\right) \, , \,
\ee
where the hierarchy conditions {\bf h1)}-{\bf h4)} mean 
$\lambda_{33}>>\lambda_{23}>>{\lambda^2_{23}/\lambda_{33}}>>\lambda_{12}$.

To leading order the eigenvalues and mixing angles are
$\lambda_3 \approx  \lambda_{33}$,
$\lambda_2  \approx  \lambda^2_{23}/\lambda_3$,
$\lambda_1 \approx \lambda^2_{12}/\lambda_2$,
$s_{23} \approx \lambda_{23}/\lambda_3$,
$s_{13} \approx \lambda_{12}s_{23}/ \lambda_3$ and
$s_{12} \approx \lambda_{12}/\lambda_2$.
\footnote{For this case exact relations have been obtained in
\cite{davi84,hara96}.}
%\ba
%\lambda_3 & \approx & \lambda_{33} \nonumber\\
%\lambda_2 & \approx & \lambda^2_{23}/\lambda_3 \nonumber\\
%\lambda_1 & \approx & \lambda^2_{12}/\lambda_2 \nonumber\\
%{s_{23} \over c_{23}} & \approx & { {\lambda_{23}} \over {\lambda_3}}
%\nonumber\\
%{s_{13} \over c_{13}} & \approx & { {\lambda_{12}s_{23}} 
%\over {\lambda_3} }
%\nonumber\\
%{s_{12} \over c_{12}} & \approx & { {\lambda_{12}} 
%\over { \lambda_2 }} \,
%\ea
We get the predictive relations 
\ba
s_{23} & \approx & \sqrt{\lambda_2 \over \lambda_3} \, \, , \nonumber\\
s_{12} & \approx & \sqrt{\lambda_1 \over \lambda_2} \, \, , \nonumber\\ 
s_{13} & \approx &s_{23} s_{12} {\lambda_2 \over \lambda_3}
\approx \sqrt{\lambda_1 \over \lambda_3} {\lambda_2 \over \lambda_3} \, \,
.
\ea

{\bf Example II:} 
\be
{\bf h} =
\left(
\begin{array}{ccc}
0 & \lambda_{12} & 0 \\
\lambda_{12} & \lambda_{22} & \lambda_{23} \\
0 & \lambda_{23} & \lambda_{33} 
\end{array}
\right) \, , \,
\ee
where in addition I will assume 
\be
\lambda_{22} \simeq \lambda_{23} \, .
\ee
Then the hierarchy conditions {\bf h1)}-{\bf h4)} give
$\lambda_{33}>>\lambda_{22},\lambda_{23}>>\lambda_{12}$.

To leading order the eigenvalues and mixing angles are
$\lambda_3 \approx  \lambda_{33}$,
$\lambda_2 \approx \lambda_{22}$,
$\lambda_1 \approx \lambda^2_{12}/\lambda_2$,
$s_{23} \approx \lambda_{23} / \lambda_3$,
$s_{13} \approx \lambda_{12}s_{23} / \lambda_3$ and
$s_{12} \approx \lambda_{12} / \lambda_2$.
%\ba
%\lambda_3 & \approx & \lambda_{33} \nonumber\\
%\lambda_2 & \approx & \lambda_{22} \nonumber\\
%\lambda_1 & \approx & \lambda^2_{12}/\lambda_2 \nonumber\\
%{s_{23} \over c_{23}} & \approx & { {\lambda_{23}} \over {\lambda_3}}
%\nonumber\\
%{s_{13} \over c_{13}} & \approx & { {\lambda_{12}s_{23}} 
%\over {\lambda_3} }
%\nonumber\\
%{s_{12} \over c_{12}} & \approx & { {\lambda_{12}} 
%\over { \lambda_2 }} \,
%\ea
Predictive relations are 
\ba
s_{12} & \approx & \sqrt{\lambda_1 \over \lambda_2} \, \, , \nonumber\\
s_{23} & \simeq & {\lambda_2 \over \lambda_3} \, \, , \nonumber\\
s_{13} & \approx & s_{23} s_{12} {\lambda_2 \over \lambda_3}
\approx \sqrt{\lambda_1 \over \lambda_2} ({\lambda_2 \over \lambda_3})^2
\, \, .
\ea
The second relation follows because I assumed $\lambda_{22} 
\simeq \lambda_{23}$. 

A note about phases: in this example all phases cannot be completely eliminated
by redefinitions of the fields. One phase will be included in the
diagonalization
matrix (see for example \cite{babu92}. For general treatment of phases see
for example \cite{kuse94,rasi97}.).

{\bf Example III:} 
\be
{\bf h} =
\left(
\begin{array}{ccc}
0 & 0 & \lambda_{13} \\
0 & 0 & \lambda_{23} \\
\lambda_{13} & \lambda_{23} & \lambda_{33} 
\end{array}
\right) \, , \,
\ee
Then the hierarchy conditions {\bf h1)}-{\bf h4)} imply
$\lambda_{33}>>\lambda_{23}>>\lambda_{13}$. 
The structure in this example appears
in references \cite{albr89,nill89,shro92}. A similar structure appears in
\cite{ramo93},
with a nonzero $\lambda_{22}$, but small (of the
order of $\lambda^2_{23} / \lambda_{33}$), so that the results and
predictions are same as here.

To leading order the eigenvalues and mixing angles are
$ \lambda_3 \approx \lambda_{33}$,
$ \lambda_2 \approx \lambda^2_{23} / \lambda_3$,
$ \lambda_1 \approx 0$,
$s_{23} \approx \lambda_{23} / \lambda_3$,
$s_{13} \approx \lambda_{13} / \lambda_3$ and
$s_{12} \approx - \lambda_{13} / \lambda_{23}$.
It is interesting that here even though the structure has a (1,3) element,
but vanishing (1,2) element, still the (1,2) rotation is bigger than the 
(1,3) rotation, as in the previous examples. 

Predictive relations are
\ba
s_{23} \approx \sqrt{\lambda_2 \over \lambda_3} \, \, ,\nonumber\\ 
s_{13} \approx s_{23} s_{12} \, \, .
\ea

{\bf Example IV:} 
\be
{\bf h} =
\left(
\begin{array}{ccc}
0 & \lambda_{12} & \lambda_{13} \\
\lambda_{12} & 0 & \lambda_{23} \\
\lambda_{13} & \lambda_{23} & \lambda_{33} 
\end{array}
\right) \, , \,
\ee
where in addition I will assume
\be
\lambda_{12} \simeq {{\lambda_{13} \lambda_{23}} \over \lambda_{33}} \, \,
, 
\ee
that is $\lambda_{12}$ still {\it smaller} then $\lambda_{13}$.
The hierarchy conditions {\bf h1)}-{\bf h4)} imply
$\lambda_{33}>>\lambda_{23}>>\lambda_{13}$
and $\lambda^2_{23}/\lambda_{33}>>\lambda_{12}$.
A similar structure appears in \cite{albr89} with various relative sizes of 
$\lambda_{13}$ and $\lambda_{12}$.

To leading order the eigenvalues and mixing angles are
$ \lambda_3 \approx \lambda_{33}$,
$ \lambda_2 \approx \lambda^2_{23} / \lambda_3$,
$ \lambda_1 \simeq \lambda^2_{12} / \lambda_2 \simeq \lambda^2_{13} /
\lambda_3 $,
$s_{23} \approx \lambda_{23} / \lambda_3$,
$s_{13} \approx \lambda_{13} / \lambda_3$ and
$s_{12} \simeq \lambda_{12} / \lambda_2$.
Still the (1,2) rotation is bigger than the 
(1,3) rotation, as in the previous examples. 

Predictive relations are
\ba
s_{23} & \approx & \sqrt{\lambda_2 \over \lambda_3} \, \, , \nonumber\\ 
s_{12} & \simeq & \sqrt{\lambda_1 \over  \lambda_2} \, \, , \nonumber\\ 
s_{13} & \simeq & s_{23} s_{12} \simeq \sqrt{\lambda_1 \over \lambda_3} 
\, \, .
\ea

It is interesting to note that the matrices in Examples III and IV
can be obtained in a U(2) flavor theory in a similar manner to Ref. \cite{barb96}.
Before the flavor symmetry is broken the only allowed
term is $\lambda_{33}$ and it is of order one.
Other elements get generated from higher dimensional operators when the
flavor
symmetry is broken down. Which of the elements get created depends now on
the flavon content on the theory. For example, one doublet
can create $\lambda_{23}$ when U(2) is broken first to U(1) 
and then $\lambda_{13}$ and $\lambda_{12}$ get created with another doublet
and an antisymmetric singlet when U(1) is broken to nothing at a lower scale.

{\bf Example V:}\cite{bran90}
\be
{\bf h} =
\left(
\begin{array}{ccc}
0 & c & 0 \\
c' & 0 & b \\
0 & b' & a 
\end{array}
\right) \, , \,
\ee
with $a >> b,b'$ and $(bb'/a) >> c,c'$.
This is an asymmetric matrix and we must diagonalize ${\bf h}^\dagger {\bf h}$.
To leading order the eigenvalues of {\bf h} and mixing angles are
\ba
m_3 = \sqrt{\lambda_3} & \approx & a 
\, \, , \nonumber\\
m_2 = \sqrt{\lambda_2} & \approx &{{b b'} \over a}
\, \, , \nonumber\\
m_1 = \sqrt{\lambda_1} & \approx & {{c c'} \over m_2} 
\, \, , \nonumber\\
s_{23} & \approx & {{b'} \over a}
\, \, , \nonumber\\
s_{13} & \approx & {{c' b} \over a^2}
\, \, , \nonumber\\
s_{12} & \approx & - {{c'} \over {({{b b'} \over a})}} \, \, .
\ea
with one relation
\be
s_{12} \approx - s_{13} s_{23} {m^2_3 \over m^2_2} \, \, .
\ee

\vspace{0.2in}

{\bf D. Is it important to do the (2,3) rotation before the (1,2)
rotation?}\hspace{0.5cm} 

As we saw in the previous section, diagonalization of hierarchical
structures is done to leading order with only (2,3) and (1,2) 
rotations with the diagonalizing matrix (\ref{leftv})
\be
{\bf R} \approx {\bf R}_{23} {\bf R}_{12} \, .
\ee

It is interesting, if not surprising, that diagonalization of the hierarchical
structures can be done to leading order in the reverse order of rotations, that is
first the (1,2) rotation, and then the (2,3) rotation. To show this let us
consider the exact results for
the diagonalizing unitary transformation 

\be 
{\bf R} = {\bf R}_{12} {\bf R}_{13} {\bf R}_{23} \, \, .
\ee

For this choice of rotations one can obtain the exact results for diagonalizing
angles similar to the case discussed in Section B,
\ba
{s_{12} \over c_{12}} & = & { {(\lambda_{33} -
\lambda_1)\lambda_{12} - \lambda_{13}\lambda_{23}} 
\over {(\lambda_{33}-\lambda_1)(\lambda_{22}-\lambda_1)-\lambda^2_{23}}}
\,\, ,
\nonumber\\
{s_{13} \over c_{13}} & = & { {\lambda_{13}c_{12} - \lambda_{23}s_{12}} 
\over {\lambda_{33}-\lambda_1} }
\,\, ,
\nonumber\\
{s_{23} \over c_{23}} & = & { {\lambda_{13}s_{12} + 
\lambda_{23} c_{12} } 
\over {(\lambda_{33}-\lambda_2)c_{13} + 
(\lambda_{13}c_{12} + \lambda_{23}s_{12})s_{13}}} \, . 
\label{exact2}
\ea
For hierarchical Yukawa structures, conditions {\bf h1)}-{\bf h4)} give  
\ba
{s_{12} \over c_{12}} & \approx & { {\lambda_{12} - 
\lambda_{13}{\lambda_{23} \over \lambda_{33}}} 
\over {\lambda_{22}-{\lambda^2_{23} \over \lambda_{33}}}} \, ,
\nonumber\\
{s_{13} \over c_{13}} & \approx & { {\lambda_{13} - \lambda_{23}s_{12}} 
\over {\lambda_{33}} } \, ,
\nonumber\\
{s_{23} \over c_{23}} & \approx & { {\lambda_{23}} 
\over {\lambda_3}} \, . 
\ea
Comparing with the approximate angles (\ref{app32}), we see that the 
rotation angles (1,2) and (2,3) agree to leading order. Only the small (1,3) 
rotation changes.

\vspace{0.2in}

{\bf E. CKM Parametrizations for the hierarchical structures}

The most general CKM
matrix can be written as a function of three angles and one
phase. Various parametrizations of CKM exist
today\cite{koba73,maia76,maia77,chau84,dimo92,pdgr96}
in which these three angles and one phase appear in various places.
It was noticed some time ago\cite{jarl89} that there are essentially
twelve different parametrizations\cite{rasi97,frit97a}, which correspond to various
ways of combining the three rotation angles in a particular parametrization.  
For each of the combinations there is a continuum of possibilities,
depending on positioning of the one nontrivial CP violating phase in the
parametrization.

Physics of the Standard Model clearly does not depend on which parametrization
we use. However, if one goes beyond the Standard Model, it might turn out more
practical to use a certain parametrization. In such a parametrization a particular
prediction, such as a relation between CKM elements and quark masses, maybe be more
transparent.

As was shown in reference \cite{rasi97}, it is always possible to get
any of the 12 possible parametrizations of the CKM matrix from any
parametrizations of the unitary matrices that diagonalize up 
and down quark masses. However, such procedure may be quite complicated,
and, in the process, possible relations between quark masses and CKM
matrix elements may be lost. Only a clever choice of a particular 
parametrization may reveal clearly such predictions, and we discuss which
one should be used in the case of hierarchical structures.

We defined hierarchical structures in the previous sections as the ones
in which the diagonalizing angles $s_{23}$ and $s_{12}$ are much bigger
than the third angle $s_{13}$. In order to discuss which CKM
parametrization to use, we need to know exactly how much bigger they are
since we need to estimate also the smallest elements
$V_{ub}$ and $V_{td}$, which will involve both $s_{13}$ and products
$s_{12}s_{23}$. We now discuss separately the relative
sizes of these two elements

{\bf Case I: $s_{13}$ rotations affecting $V_{ub}$ or $V_{td}$}

If $s_{13}$ is of the order of $s_{12}s_{23}$ we cannot
neglect this rotation when estimating $V_{ub}$ and 
$V_{td}$. In this case, in the most general case
when both up and down quark mass matrices need to be diagonalized
with three rotations each, the analysis is quite complicated 
and one has to resort to the exact results \cite{rasi97}.
However, in the case of simpler structures where one of
the quark mass matrices is diagonalized with only one rotation,
there are preffered CKM parametrizations.

Suppose that the down quark mass matrix has only mixing between
the first two families, for example of the
Fritzsch-Weinberg-Wilczek-Zee type \cite{frit77,fwzw}
\be
{\bf h}^d =
\left(
\begin{array}{ccc}
0 & F & 0 \\
F & E & 0 \\
0 & 0 & D 
\end{array}
\right) \, , \,
\label{wz}
\ee
so that the diagonalizing matrix is
\be
{\bf V}^d = {\bf R}_{12},
\ee 
with 
\be
s^d_{12} \approx \sqrt{m_d / m_s}.
\label{pred1}
\ee 
Then we should choose the three
rotations that diagonalize the up quark matrix such that the first
rotation is a $s^u_{12}$ rotation, so that it is trivially combined 
with the $s^d_{12}$ into a single $s_{12}$. Of the other two rotations in
the up diagonalizing matrix, one should be a $s_{23}$ rotation (since
we assume hierarchical matrices), but the choice for the last rotation and 
the order of rotations should be chosen only by the criteria of
predictivity. Thus, for this case possible parametrizations are
\be
{\bf V} = 
{\bf R}_{12} {\bf R}_{23} {\bf R}_{12} \, , \, 
{\bf R}_{13} {\bf R}_{23} {\bf R}_{12} \, , \, 
{\bf R}_{23} {\bf R}_{13} {\bf R}_{12} \, . \, 
\ee
As an example let us assume that, in addition to the form (\ref{wz}) for the
down quark mass matrix, the up type quark mass matrix is of the
form given in Example IV
\be
{\bf h}^u =
\left(
\begin{array}{ccc}
0 & C' & C \\
C' & 0 & B \\
C & B & A 
\end{array}
\right) \, , \,
\ee
where $C' \simeq C B / A$. Here $s_{13}$ is exactly of the order
$s_{12}s_{23}$ and it needs to be included in the CKM matrix. Thus a
choice for the up quark diagonalizing matrix is
\be 
{\bf R}^u = {\bf R}^u_{12} {\bf R}^u_{13} {\bf R}^u_{23} \, \, ,
\ee
where, to leading order, (see Example IV) 
\be
s^u_{12} \simeq \sqrt{m_u \over m_c} \, , \,
s^u_{23} \approx \sqrt{m_c \over m_t} \, , \, 
s^u_{13} \approx \sqrt{m_u \over m_t}.
\label{pred2}
\ee
The CKM matrix that one obtains is
\be
{\bf V} = R^{u\dagger}R^d = {\bf R}^{uT}_{23} {\bf R}^{uT}_{13} {\bf
R}_{12} \, \, ,
\ee
where $s_{12} = s^d_{12} - s^u_{12}$. This is the ``standard CKM parametrization"
of Chau, Keung and Maiani \cite{chau84,maia76}. With the above predictions
(\ref{pred1}) and (\ref{pred2}) the CKM elements are successfully
reproduced\footnote{I am interested here only in approximate relations.
These relations are
successful within a factor of 2 or 3, which can easily be accommodated in the
original Yukawa matrices by numerical factors of order one.}. A clear
prediction here is 
\be
|{V_{ub} \over V_{cb}}| \approx {s^u_{13} \over s^u_{23}} \simeq \sqrt{m_u
\over m_c} \, \, .
\ee

\newpage

{\bf Case II: $s_{13}$ rotations too small to affect $V_{ub}$ or $V_{td}$
to leading order}

In order for the angles $s^u_{13}$ and $s^d_{13}$ not to contribute
to $V_{ub}$ and $V_{cb}$ to leading order, following conditions
must be satisfied\cite{hall93}
\be
|s^d_{13}-s^u_{13}| << s^u_{12} |s^d_{23} - s^u_{23}| \, ; \,
|s^d_{13}-s^u_{13}| << s^d_{12} |s^d_{23} - s^u_{23}| \, \, .
\ee
The above relations are the exact conditions on $\lambda^u_{13}$ and
$\lambda^d_{13}$, but are a bit complicated to write down explicitly
in terms of $\lambda_{ij}$. 
As a guideline, a somewhat less restricting, but more
understandable condition $s_{13} << s_{12}s_{23}$ is obtained when
\be
\lambda_{13} << \lambda_{12} s_{23} \, \, ,
\ee
in both up and down sectors.

Thus in this case we need to consider only the $s_{23}$ and $s_{12}$ rotations.
As was shown in Section D, the values of these two angles do not depend on
the order in which they are taken when diagonalizing a quark mass matrix.
However, depending on 
the order two simple CKM parametrizations emerge, and we discuss them next
in detail.  If one diagonalizes up and down quark matrices with first
(2,3) rotations and
then (1,2) rotations, the CKM matrix is
\be
{\bf V} = R^{u\dagger} R^d \approx
R^{uT}_{12}R_{23}R^d_{12} \, \, ,
\label{ckm1}
\ee
where 
$\theta_{23} = \theta^d_{23} - \theta^u_{23}$.
This is a parametrization of the CKM in terms of three 
rotation angles. 
What is 
nice is that these angles are directly related to angles of the original
diagonalizing matrices. Since one
can write an {\it exact} parametrization in terms of three rotation angles, the
exact
angles will differ from the above angles only by small corrections that are
subleading to the (1,2) and (2,3) rotations that we used.

If one allows Yukawa matrices to be complex, one can show \cite{rasi97} that it
amounts to putting one complex phase $\delta$ in the CKM (\ref{ckm1}) between the
rotations.
Also, the (2,3) rotation will now in general be only the absolute value of
the sum of the (2,3) rotations with a relative complex phase
\be
\theta_{23} = |\theta^d_{23} - e^{i\alpha}\theta^u_{23}|.
\ee
so that the relation of that CKM angle with the original diagonalizing 
(2,3) angles gets blurred. However, to leading order, the (1,2) CKM angles
{\it are} the (1,2) angles that diagonalize the up and down sector.
For completeness, let us list a complex CKM generalization of (\ref{ckm1}) 
\footnote{For any product of three angles there is a continuum of possibilities
for placement of the phase $\delta$.}
\ba
{\bf V} & = &
{\rm diag}(e^{-i\delta}11){\bf R}^{Tu}_{12}{\rm diag}(e^{i\delta}11){\bf
R}_{23}{\bf R}^d_{12} \nonumber\\
& = &
\left(
\begin{array}{ccc}
c^u_{12} c^d_{12} + s^d_{12} s^u_{12} c_{23} e^{-i\delta} &
c^u_{12} s^d_{12} - s^u_{12} c_{23} c^d_{12} e^{-i\delta}   &
-s^u_{12} s_{23} e^{-i\delta} \\
 s^u_{12} c^d_{12} e^{i\delta} - s^d_{12} c^u_{12} c_{23} &
+s^u_{12} s^d_{12} e^{i\delta} + c^u_{12} c^d_{12} c_{23}  &
c^u_{12} s_{23} \\ 
s^d_{12} s_{23} &
- c^d_{12} s_{23} &
c_{23}
\end{array}
\right) \, .
\label{dhr}
\ea
This parametrization appears in  
\cite{dimo92,hall93,ande93,barb96,frit97}. 
One can immediately write the following relations
\be
|{V_{ub} \over V_{cb}}| = {s^u_{12} \over c^u_{12}} \, \, ; \, \,
|{V_{td} \over V_{ts}}| = {s^d_{12} \over c^d_{12}} \, .
\label{rels1}
\ee

Similarly, if one diagonalizes up and down quark matrices 
with first (1,2) rotations and
then (2,3) rotations, the CKM matrix is
\be
{\bf V} = R^{u\dagger} R^d \approx
R^{uT}_{23}R_{12}R^d_{23} \, ,
\label{ckm2}
\ee
where $\theta_{12} = \theta^d_{12} - \theta^u_{12}$, so that the parametrization is
again given in terms of three angles. The complex generalization can be written
as
\ba
{\bf V} & = &
{\rm diag}(11e^{-i\delta}){\bf R}^{uT}_{23}
{\rm diag}(11e^{i\delta}){\bf R}_{12}{\bf R}^d_{23} \nonumber\\
& = &
\left(
\begin{array}{ccc}
c_{12} &
s_{12} c^d_{23} &
s_{12} s^d_{23} \\
- s_{12} c^u_{23} &
c_{12} c^u_{23} c^d_{23} + s^u_{23} s^d_{23} e^{i\delta} &
c_{12} c^u_{23} s^d_{23} - s^u_{23} c^d_{23} e^{i\delta} \\
-s_{12} s^u_{23} e^{-i\delta} &
c_{12} c^d_{23} s^u_{23} e^{-i\delta} - s^d_{23} c^u_{23} &
c_{12} s^d_{23} s^u_{23} e^{-i\delta} + c^u_{23} c^d_{23}
\end{array}
\right)
\label{km}
\ea
It is the parametrization originally proposed 
by Kobayashi and Maskawa\cite{koba73}.
It has the following predictions
\be
|{V_{td} \over V_{cd}}| = {s^u_{23} \over c^u_{23}} \, \, ; \, \,
|{V_{ub} \over V_{us}}| = {s^d_{23} \over c^d_{23}} \, .
\label{rels2}
\ee

A final note. The parametrization (\ref{km}) was discarded in reference
\cite{frit97a} on two grounds which I now want to argue to be
unnecessary. Let us
first assume there is no CP
violating phase. The first requirement that there be only
one ${\bf R}_{23}$ rotation in a particular parametrization because there should
be just one angle when the first generation decouples is too strong, since
in the parametrization (\ref{km}) when the ${\bf R}_{12}$ approaches unity, the
two ${\bf R}_{23}$ rotations trivially combine into a single one. Even with the
CP violating phase between two ${\bf R}_{23}$ rotations, it is easy to write it
as one ${\bf R}_{23}$ rotation between two phase transformations (see for
example
Ref. \cite{rasi97}). The second requirement that the CP violating phase
should disappear when the first generation masses disappear is also too
strong. What one should ask from a parametrization is that the phase
disappears when the mixing between a certain generation with the other two
generations disappears. From this
standpoint both parametrizations (\ref{dhr}) and (\ref{km}) are
acceptable: in parametrization (\ref{dhr}) the phase disappears when the
first generation does not mix with the rest (${\bf R}_{12}={\bf 1}$), and
similarly in parametrization (\ref{km}) when the mixings with the third
generation vanish (${\bf R}_{23}={\bf 1}$).

\vspace{0.2in}

{\bf F. Discussion and Conclusions}

In conclusion, I have defined general hierarchical structures of the Yukawa
matrices with the four conditions {\bf h1)}-{\bf h4)}. I defined them as 
structures with hierarchy in Yukawa elements and their
eigenvalues, and by demanding that the (1,3) rotation be much smaller than the
(1,2) and (2,3) rotations. Then such structures can be diagonalized with
the same (1,2) and (2,3) rotations in any order. 

In addition to the more familiar Fritzsch structures, 
examples of such structures
include matrices which have nonnegligible (1,3) elements.
They can be used to describe quark masses and mixings and
examples of such matrices were given in Section C (Examples III and IV.).
It is interesting to note that it is possible to build such matrices in a
U(2) flavor theory similar to Reference \cite{barb96}, with an additional
doublet flavon\cite{rasi98}.

We studied the diagonalization of hierarchical quark
mass matrices and the CKM parametrizations that naturally follow from such
matrices. When the $s_{13}$ contributions to $V_{ub}$ and 
$V_{td}$ cannot be neglected, any CKM parametrization with at least one
(2,3) and
one (1,2) rotation may appear useful, that is it will depend on the underlying
flavor theory which of the CKM parametrizations will be most transparent to
predictions of the theory.
 
If, further, the (1,3) rotations can be
neglected in the CKM, two possible
CKM parametrizations that simply relate the CKM elements to the diagonalizing
angles appear, namely ${\bf R}_{12}{\bf R}_{23}{\bf R}_{12}$ and 
${\bf R}_{23}{\bf R}_{12}{\bf R}_{23}$. 
Which one of the two parametrizations should one use? This will
depend on the underlying flavor theory. If the theory has nice predictions
for the (1,2) rotations in terms of quark masses one should use the
parametrization (\ref{ckm1}). Examples of this type of theories include
generalized Fritzsch structures (examples I and II in Section C)
where $s^d_{12} \approx \sqrt{m_d/m_s}$ and $s^u_{12} \approx
\sqrt{m_u/m_c}$. 
In this case, we get the clear predictions
\cite{frit79,stec83,dimo92,hall93}
\be
|{V_{ub} \over V_{cb}}| \approx \sqrt{m_u \over m_c} \, \, ; \, \,
|{V_{td} \over V_{ts}}| \approx \sqrt{m_d \over m_s} \, .
\ee
However, one can also have flavor theories of the second and third generation masses
where, with first generation masses being small and their estimation not so
reliable. Then the second case might be more applicable if there are clear
predictions of the (2,3) rotations in terms of quark masses. For example
if the up quark matrix is of the type shown in Example I, and down quark
matrix of type II we get relations\cite{barb96,rasi97a}
$s^d_{23} \simeq m_s/m_b$ and $s^u_{23} \approx \sqrt{m_c/m_t}$. The predictions
are then
visible in Case II where
\be
|{V_{td} \over V_{cd}}| \approx \sqrt{m_c \over m_t} \, \, ; \, \,
|{V_{ub} \over V_{us}}| \simeq {m_s \over m_b} \, .
\ee

Of course, physics of the Standard Model does not depend on which CKM
parametrization one uses. However, if there is a flavor theory, as we strongly
believe, it will hopefully reduce the number of parameters
and produce some predictions. Then, it is important to have clear and simple
formulas for the predictions, and this depends on which
parametrizations one uses. Which one of the two parametrizations one should use
depends on the underlying flavor theory, i.e. on which diagonalizing angles one
can relate to the quark masses. 

\vspace{0.2in}

{\bf G. Acknowledgments}

I would like to thank K.S. Babu and Atsushi Yamada for valuable comments, as well as
Lawrence Hall and the LBNL theory group for hospitality during my stay in Berkeley. 
Part of this work was done during  the CPMASS97 School and Workshop
in Lisbon and the ICTP Extended Workshop on Highlights in Astroparticle Physics.
This work is partially supported by EEC grant under the TMR contract
ERBFMRX-CT960090.

\end{document}